\documentclass[aps,twocolumn,showpacs,preprintnumbers,amsmath,amssymb]{revtex4}
\usepackage{graphicx}
\usepackage{color}
\definecolor{navyblue}{rgb}{0.3,0.3,1}
\definecolor{purple}{rgb}{0.6,0,0.5}
\usepackage[colorlinks=true, pdfstartview=FitV, linkcolor=purple, citecolor=
blue,urlcolor=navyblue]{hyperref}

\begin{document}

\title{Polyakov-Nambu-Jona-Lasinio phase diagrams and quarkyonic phase from order
parameters}
\author{M. Dutra$^1$, O. Louren\c co$^2$, A. Delfino$^3$, T. Frederico$^1$,
and M. Malheiro$^1$}

\affiliation{$^1$Departamento de F\'isica, Instituto Tecnol\'ogico de
Aeron\'autica, DCTA, 12.228-900, S\~ao Jos\'e dos Campos, SP, Brazil\\
$^2$Departamento de Ci\^encias da Natureza, Matem\'atica e Educa\c
c\~ao, CCA, Universidade Federal de S\~ao Carlos, 13600-970, Araras, SP, Brazil\\
$^3$Instituto de F\'isica - Universidade Federal Fluminense,
Av. Litor\^ anea s/n, 24210-150 Boa Viagem, Niter\'oi RJ, Brazil}

\date{\today}

\begin{abstract}
We show that the magnitude of the order parameters in Polyakov-Nambu-Jona-Lasinio (PNJL)
model, given by the quark condensate and the Polyakov loop, can be used as a criterium to
clearly identify, without ambiguities, phases and boundaries of the strongly interacting
matter, namely, the broken/restored chiral symmetry, and confinement/deconfinement
regions. This structure is represented by the projection of the order parameters in the
temperature-chemical potential plane, which allows a clear identification of  pattern
changes in the phase diagram. Such a criterium also enables the emergence of a quarkyonic
phase even in the two-flavor system. We still show that this new phase diminishes
due to the influence of an additional vector-type interaction in the PNJL phase diagrams,
and is quite sensitive to the effect of the change of the $T_0$ parameter in the Polyakov
potential. Finally, we show that the phases and boundaries constructed by our method
indicate that the order parameters should be more strongly correlated, as in the case of
entanglement PNJL (EPNJL) model. This result suggests a novel way to pursue further
investigation of new interactions between the order parameters in order to improve the
PNJL model.
\end{abstract}

\maketitle

\section{Introduction}

In the large distances, or equivalently, low energies regime, one of the methods to
treat Quantum Chromodynamics (QCD) is the numerical lattice calculations~\cite{lattice} based
on Monte Carlo simulations~\cite{mcarlo}. The results from these techniques are
provided for the pure gluon sector, i.e., in the limit of infinitely heavy quarks, as well
as for systems including dynamical quarks. The latter systems, however, face the fermion
sign problem~\cite{sign} at finite quark chemical potential ($\mu_q$) regime.
Nevertheless, such a problem is circumvented by reweighting methods, density of state ones
and other, see Refs.~\mbox{\cite{expansao,reweighting,review,imaginary,dos}} for such
treatments. Different from this method, other approach to describe QCD is the use
of effective models such as the MIT bag model~\cite{mit} and the Nambu-Jona-Lasinio (NJL)
one~{\mbox{\cite{njl,buballa,outros}}. The former treats gluons and massless quarks as free
particles in which the confinement phenomenon is incorporated by including a bag constant in an
{\it ad hoc} fashion.  The latter presents further similarities with the full QCD
theory but does not take into account the confinement, since quarks interact each
other via pointlike interactions without exchanged gluons.

In order to become the NJL model still more realistic, taking into account the
quark confinement at low energies, Fukushima~\cite{fuku1} developed the named
Polyakov-Nambu-Jona-Lasinio model (PNJL), in which the confinement is included in the NJL
structure through the Polyakov loop $\Phi=e^{-F_q/T}$, where $F_{q}$ stands for the quark
free-energy (in Ref.~\cite{arriola}, it is argued that $\Phi$ can be also represented by
hadronic states). From this widely studied effective QCD
model~\cite{bhattacharyya,kouno,rgatto,weise1,weise2,weise4,weise6,ratti,costa,scoccola,
fuku2,blanquier,hass,blaschke}, many informations on the strongly interacting matter can
be obtained, such as its phase diagram~\cite{fukupd}, where the proper broken/restored
chiral symmetry, and confinement/deconfinement regions are identified. Other typical
approaches are the use of two equations of state~\cite{twoeos} in the description of the
quark phase and the hadronic one~\cite{beguntawfik}, as well as hybrid
models~\cite{hybrid}. Moreover, other effective models coupled to the Polyakov loop are
equally useful~\cite{pqm,ueda,vivek,tkahara,marko,mao}. 

Different ways, based on different criteria, to construct the phase diagram in the
$T\times\mu_q$ plane are addressed in the literature. In this work we compare such
criteria and present a new one in order to clearly identify the regions and boundaries of the
quark phase diagram, generated exclusively from the PNJL model. Our analysis suggests
that the order parameters should be correlated as in the entanglement PNJL
model, if the coincidence seen for the chiral and confinement transitions
obtained from lattice QCD calculations at high T and very small $\mu_q$ is also confirmed
for small temperatures and larger quark densities values. This investigation follows a
sequence of studies presented by our group in previous works~\cite{prd1,prd2}, all of them
motivated by the search of a better description of phases and boundaries of strongly
interacting matter, specifically through analysis of PNJL phase diagrams. 

The regime of high $\mu_q$ and very low temperatures is very important to investigate 
the existence o quark matter in the core of neutron stars or even in bare quark
stars~\cite{qstars}, one of the most important questions nowadays concerning the
internal matter composition of compact stars, in particular if the quarkyonic phase is
presented or not. Therefore, our findings are useful, for instance, in the study of
protoneutron stars that are described at $T<50$~MeV. Applications of PNJL models to
compact stars have been done recently for the protoneutron stars
evolution~\cite{shao,lugones}, for quark~\cite{coelho}, and for hybrid
stars~\cite{hybridstars}. Furthermore, we also point out that investigations of the quark
phase diagram are relevant for a deeper understanding of the strongly interacting matter.
The predictions of such studies, specially at the high density regime, will be tested in
future experiments~\cite{fair,nica}.

The paper is organized as follows. In Sec.~\ref{pnjlmodel} the basic thermodynamical
quantities of the PNJL model are presented and also the distinct Polyakov
potentials used in the literature. In Sec.~\ref{discussion} we discuss the PNJL phase
diagrams constructed from different criteria, and present our new method based on the
magnitude of the order parameters. We also discuss the effect of the repulsive interaction
in the PNJL phase diagrams obtained from our method. The change of an specific parameter
in the Polyakov potential of the PNJL model is analyzed in this section, as well as the
EPNJL model. Finally, in Sec.~\ref{summary} the summary and our mainly conclusions are
showed.

\section{PNJL model}
\label{pnjlmodel}

The connection between the fermion ($q$) and the gauge ($A^\mu$) field in the PNJL model
is achieved by making the substitution $\partial^{\mu}\rightarrow D^\mu=\partial^{\mu} -
iA^{\mu}$ in the Lagrangian density, where $A^\mu=\delta^\mu_0A_0$ and
$A_0=gA_\alpha^0\frac{\lambda_\alpha}{2}$ ($g$ is the gauge coupling and $\lambda_\alpha
$ are the Gell-Mann matrices). Techniques from field theory at finite temperature, as
those used in Ref.~\cite{weise2}, are applied to get the 
following grand canonical potential per volume,
\begin{eqnarray}
\Omega_{\mbox{\tiny PNJL}} &=& \mathcal{U}(\Phi,\Phi^*,T) +
G_s\rho_s^2 - \frac{\gamma}{2\pi^2}\int_0^{\Lambda}E\,k^2dk
\nonumber \\
&-& \frac{\gamma}{6\pi^2}\int_0^{\infty}\frac{k^4}{E}dk
\left[ F(E,T,\mu_q,\Phi,\Phi^*) \right.\nonumber \\
&+& \left. \bar{F}(E,T,\mu_q,\Phi,\Phi^*) \right]
\label{omega}
\end{eqnarray}
where $E=E(M)=(k^2+M^2)^{1/2}$, $\rho_s$ is the quark condensate given
by $\rho_s=\left<\bar{q}q\right>=\left<\bar{u}u\right> +
\left<\bar{d}d\right>=2\left<\bar{u}u\right>$ in the isospin symmetric system,
and $\gamma=N_s\times N_f\times N_c=12$ is the degeneracy factor due to the spin
($N_s=2$), flavor ($N_f=2$), and color numbers ($N_c=3$). The constituent quark
mass is $M=m_0-2G_s\rho_s$. The second integral in Eq.~(\ref{omega}) leads to the
expected Stefan-Boltzmann limit, since the momentum of the active quarks are
unconstrained.

The traced Polyakov loop is defined in terms of $A_4=iA_0\equiv T\phi$ as
\begin{align}
\Phi&=\frac{1}{3}\rm{Tr}\left[\,\,\rm{exp}\left(i\int_0^{1/T}d\tau\,A_4\right)\right]
\nonumber \\
&=\frac{1}{3}\rm{Tr}\left[\rm{exp}(i\phi)\right]
=\frac{1}{3}\rm{Tr}\left\lbrace\rm{exp}[i(\phi_3\lambda_3+\phi_8\lambda_8)]
\right\rbrace \nonumber \\
&=
\frac{1}{3}\left[\rm{e}^{i(\phi_3+\phi_8/\sqrt{3})}+\rm{e}^{i(-\phi_3+\phi_8/\sqrt{3})}
+\rm{e}^{-2i\phi_8/\sqrt{3}}\right],
\label{traced}
\end{align}
in a gauge (Polyakov gauge) in which the gluon field is written in terms of the diagonal
Gell-Mann matrices as $\phi=\phi_3\lambda_3+\phi_8\lambda_8$, with $\phi_3,\phi_8 \in 
\mathbb{R}$. Here, the definitions $\phi_3=A_4^3/T$ and $\phi_8=A_4^8/T$ were taken into
account. It is worth to mention that $\Phi^*$ is the complex conjugate of the
complex field~$\Phi$.

As pointed out in Refs.~\cite{ratti,costa}, an important consequence of the coupling
between $\Phi$ and the quark sector, is the possibility to deal with the PNJL model in the
same theoretical way as in the NJL one, regarding the statistical treatment. However, in
this case new distributions functions for quarks and antiquarks appear and are given by,
\begin{align}
&F(E,T,\mu_q,\Phi,\Phi^*) = \nonumber \\
&=\frac{\Phi e^{2(E-\mu_q)/T} +
2\Phi^*e^{(E-\mu_q)/T} + 1}{3\Phi e^{2(E-\mu_q)/T} + 3\Phi^*
e^{(E-\mu_q)/T} + e^{3(E-\mu_q)/T} + 1},
\label{fdmp}
\end{align}
and $\bar{F}(E,T,\mu_q,\Phi,\Phi^*) = F(E,T,-\mu_q,\Phi^*,\Phi)$, generalized from the
usual Fermi-Dirac distributions by the inclusion of $\Phi$ and $\Phi^*$. Another
difference in the PNJL model is the Polyakov loop potential $\mathcal{U}(\Phi,\Phi^*,T)$.
Some versions of this potential were proposed in the literature, and following the
language of Ref.~\cite{fuku2}, we refer two of them by RTW05~\cite{weise1}, and
RRW06~\cite{weise2,weise4,weise6,ratti,costa,scoccola}. The other
two ones are FUKU08~\cite{fuku2}, and DS10~\cite{schramm}. Their functional forms are
given, respectively, by
\begin{eqnarray}
\frac{\mathcal{U}_{\mbox{\tiny RTW05}}}{T^4} &=& -\frac{b_2(T)}{2}\Phi\Phi^*
- \frac{b_3}{6}(\Phi^3 + {\Phi^*}^3) + \frac{b_4}{4}(\Phi\Phi^*)^2, \hspace{0.28cm}
\label{rtw05}\\
\frac{\mathcal{U}_{\mbox{\tiny RRW06}}}{T^4} &=& -\frac{b_2(T)}{2}\Phi\Phi^*
+ b_4(T)\mbox{ln}\left[h(\Phi,\Phi^*)\right],
\label{rrw06} \\
\frac{\mathcal{U}_{\mbox{\tiny FUKU08}}}{b\,T} &=& -54e^{-a/T}\Phi\Phi^*
+ \mbox{ln}\left[h(\Phi,\Phi^*)\right],
\label{fuku08} \\
\mathcal{U}_{\mbox{\tiny DS10}} &=& (a_0T^4 + a_1\mu_q^4 + a_2T^2\mu_q^2)\Phi^2\nonumber\\
&+& a_3T_0^4\mbox{ln}\left[h(\Phi,\Phi)\right],
\label{ds10}
\end{eqnarray}
where
\begin{eqnarray}
b_2(T) &=& a_0 + a_1\left(\frac{T_0}{T}\right) + a_2\left(\frac{T_0}{T}\right)^2
+ a_3\left(\frac{T_0}{T}\right)^3, \quad \\
h(\Phi,\Phi^*) &=&1 - 6\Phi\Phi^* + 4(\Phi^3 + {\Phi^*}^3) -
3(\Phi\Phi^*)^2,
\end{eqnarray}
and 
\begin{eqnarray*}
b_4(T) = b_4\left(\frac{T_0}{T}\right)^3.
\end{eqnarray*}
The constants of these parametrizations are given in Table~\ref{tab1}.

In a general way, the Polyakov potentials are constructed in order to reproduce
the well established data from lattice calculations of the pure gluon sector
(where $\Phi=\Phi^*$), concerning the temperature dependence of the Polyakov
loop and its first order phase transition, characterized by the jump of
$\Phi$ from the vanishing to a finite value at $T_0=270$~MeV (see the dotted
curve of Fig. 2 in Ref.~\cite{weise2}, for instance).
\begin{table}[!ht]
\centering
\begin{ruledtabular}
\caption{Dimensionless parameters of the potentials given
in Eqs.~(\ref{rtw05})-(\ref{rrw06}) and (\ref{ds10}). The constants of the FUKU08
potential are given by $a=664$~MeV and $b=0.03\Lambda^3$~MeV$^3$.}
\begin{tabular}{l c c c c c c}
Potentials  & $a_0$   & $a_1$   & $a_2$   & $a_3$   & $b_3$  & $b_4$   \\ 
\hline
RTW05       & $6.75$  & $-1.95$ & $2.625$ & $-7.44$ & $0.75$ & $7.5$   \\
RRW06       & $3.51$  & $-2.47$ & $15.22$ & -     & -    & $-1.75$ \\
DS10	    & $-1.85$ & $-1.44\times10^{-3}$ & $-0.08$ & $-0.40$ & - & -
\end{tabular}
\label{tab1}
\end{ruledtabular}
\end{table}

The free parameters $\Lambda = 651$ MeV, $m_0 = 5.5$ MeV, and $G_s=5.04$
GeV$^{-2}$, are obtained from the NJL sector of the PNJL model in order to
reproduce the vacuum values of $m_\pi=140.51$ MeV, \mbox{$f_\pi=94.04$} MeV, and
$|\left<\bar{u}u\right>|^{1/3}=251.32$ MeV for the pion mass, the pion
decay constant, and the quark condensate, respectively.

To completely define the model, and consequently construct its phase diagram,
one needs to evaluate the order parameters. This is done by requiring that
$\Omega_{\mbox{\tiny PNJL}}$ is minimized in respect to the set of fields of the model,
i. e., ($\rho_s$, $\phi_3$, $\phi_8$), or, equivalently, \mbox{($\rho_s$, $\Phi$,
$\Phi^*$)}. Therefore, the condition
\begin{eqnarray}
\frac{\partial\Omega_{\mbox{\tiny PNJL}}}{\partial X_i} = 0
\label{extrema}
\end{eqnarray}
with $X_i=\rho_s,\phi_3,\phi_8$, or, $X_i=\rho_s,\Phi,\Phi^*$ has to be satisfied.
However, as pointed out in Ref.~\cite{mintz} in the context of the Polyakov-quark-meson
(PQM) model, Eq.~(\ref{extrema}) is only a necessary but not sufficient condition to
ensure that the values of $X_i$ minimize $\Omega_{\mbox{\tiny PNJL}}$. The authors discuss
two distinct situations in which $\Omega_{\mbox{\tiny PNJL}}$ presents no minima. The first
of them is related to the fact that $\Omega_{\mbox{\tiny PNJL}}$  is in general complex-valued
function due to the complex fields $\Phi$ and $\Phi^*$. In this case, the minimum can not be
defined. One way to circumvent this problem very often used in literature, is to make
$\Omega_{\mbox{\tiny PNJL}}$ a real function, by requiring that $\Phi$ and $\Phi^*$ be
real and independent quantities. However, this assumption itself is not sufficient to
ensure that the conditions (\ref{extrema}) provide a field configuration which minimizes
$\Omega_{\mbox{\tiny PNJL}}$. 

The authors of Ref.~\cite{mintz} show that some Polyakov potentials such as RTW05 and
RRW06 given, respectively, in Eqs.~(\ref{rtw05}) and (\ref{rrw06}), are unbounded from
below for some values of the real quantities $\Phi$ and $\Phi^*$. For instance, it was
shown that for $\Phi^*\to\infty$ and $\Phi=0$, one has $\mathcal{U}_{\mbox{\tiny
RTW05}},\,\,\mathcal{U}_{\mbox{\tiny RRW06}}\to -\infty$. Therefore, there are no minima
for $\Omega_{\mbox{\tiny PNJL}}$ in such cases, even with $\Omega_{\mbox{\tiny PNJL}}$
being a real function. In order to ensure that the real fields minimize
$\Omega_{\mbox{\tiny PNJL}}$, the authors suggest the use of condition (\ref{extrema}) 
with the additional positivity constraint of all eigenvalues of the respective $i\times i$
Hessian matrix. The use of Eq.~(\ref{extrema}) without additional constraints to find the
mean fields of the model is called saddle point approach, frequently used in literature.

In our work, we will use the mean-field approximation described in
Ref's~\cite{weise4,weise6} that takes into account the mean-field configuration in which
$\phi_8=0$ in Eq.~(\ref{traced}). In this case, $\Phi=\Phi^*=[2\cos(\phi_3)+1]/3$ even for
$\mu_q>0$, which leads to $\Omega_{\mbox{\tiny PNJL}}\in\mathbb{R}$. Another feature of
this approach is that we do not have the problem of $\mathcal{U}_{\mbox{\tiny
RTW05}},\,\,\mathcal{U}_{\mbox{\tiny RRW06}}\to-\infty$, previously raised. 

The condition given in Eq.~(\ref{extrema}), namely,
\begin{eqnarray}
\frac{\partial\Omega_{\mbox{\tiny PNJL}}}{\partial\rho_s} =
\frac{\partial\Omega_{\mbox{\tiny PNJL}}}{\partial\Phi} = 0,
\end{eqnarray}
generates the following set of coupled equations to be solved:
\begin{equation}
M - m_0 + 2G_s\rho_s[M,E(M),T,\mu_q,\Phi] = 0,
\label{mass}
\end{equation}
and
\begin{eqnarray}
\frac{\partial\mathcal{U}(\Phi,T)}{\partial\Phi} &-& \frac{3T\gamma}
{2\pi^2N_c}\int_0^{\infty}
k^2dk[g(E(M),T,\mu_q,\Phi) \nonumber \\
&+& g(E(M),T,-\mu_q,\Phi)] = 0,
\label{loop}
\end{eqnarray}
where the function $g(E,T,\mu_q,\Phi)$ leads to
\begin{align}
&g(E,T,\mu_q,\Phi) = \nonumber \\
&=\frac{1+e^{-(E-\mu_q)/T}}{3\Phi[1+e^{-(E-\mu_q)/T}] +
e^{(E-\mu_q)/T} + e^{-2(E-\mu_q)/T}}.
\end{align}
The quark condensate is given by
\begin{eqnarray}
\rho_s &=& \frac{\gamma
M}{2\pi^2}\int_0^{\infty}\frac{k^2}{E}dk
\left[ F(E,T,\mu_q,\Phi) + \bar{F}(E,T,\mu_q,\Phi) \right] \nonumber \\
&-& \frac{\gamma M}{2\pi^2}\int_0^{\Lambda}\frac{k^2}{E(M)}dk.
\label{qcond}
\end{eqnarray}

Our study is based on the  aforementioned saddle point approach. In addition, in the
case of the mean-field approximation used here ($\phi_8=0$), we have also checked the sign
of the Hessian matrix eigenvalues. In general, our solutions correspond to minima of
$\Omega_{\mbox{\tiny PNJL}}$. Negative eigenvalues were found only for a small
region of $T$ and $\mu_q$ around the first order phase transition. Nevertheless, the
projection of the order parameters in the $T\times\mu_q$ plane from the saddle point
approach, that we will present in next section, does not differ significantly from the one
obtained by the method proposed in Ref.~\cite{mintz}. Our calculations confirm the
findings of Ref.~\cite{mintz}, that the phase boundary is not changed by considering the
saddle point approach for the  RRW06  potential, adopted in our work.

It is worth noting that in such equations, we still did not consider the repulsive
interaction, which has its magnitude given by the coupling constant $G_V$. The impact
of this specific interaction in the phase diagrams will be analyzed in
Sec.~\ref{discussion-vec}, and in these cases the saddle point solutions provide
minima of $\Omega_{\mbox{\tiny PNJL}}$.

\section{Results and Discussions}
\label{discussion}

\subsection{Phase diagrams without vector interaction}

For each pair ($T,\mu_q$), the Eqs. (\ref{mass}) and (\ref{loop})
are solved for the quantities $M$, and the Polyakov loop, that are used
in Eq. (\ref{qcond}) in order to evaluate the quark condensate. In this way, one
has for the given pair ($T,\mu_q$) both order parameters, $\rho_s$, and $\Phi$,
basic thermodynamical quantities used to construct the quark phase diagram.
Frequently, many authors use the criterium of finding the maxima of
$\partial\rho_s/\partial T$ and $\partial\Phi/\partial T$ to generate the
\mbox{$T\times\mu_q$} diagram. Thus, from this assumption, one can obtain the
following behavior depicted in Fig.~\ref{derivatives}, constructed for the
RRW06 parametrization at vanishing chemical potential.
\begin{figure}[!htb]
\centering
\includegraphics[scale=0.32]{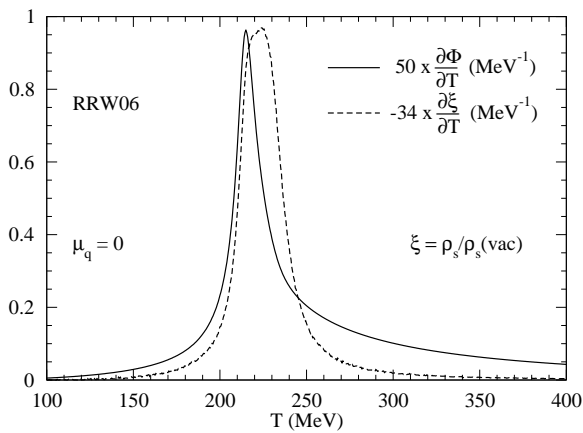}
\caption{Temperature derivatives of the order parameters as a function of $T$ for the
RRW06 parametrization at $\mu_q=0$.}
\label{derivatives}
\end{figure}

This behavior indicates a smooth crossover instead of a first order phase
transition. Therefore, the transition temperature (or pseudo-critical temperature as is
also named), is defined as that in which a maximum is found. In case of $\mu_q=0$, the
maxima of $\partial\rho_s/\partial T$ and $\partial\Phi/\partial T$ occur practically at
the same temperature, $T\sim 220$~MeV. Thus, the corresponding point in the $T\times\mu_q$
plane for this case is $\mu_q=0$, $T\sim 220$~MeV.

The peak structure of $\partial\Phi/\partial T$ exhibited in Fig.~\ref{peaks}a does not
keep the same as in Fig.~\ref{derivatives} for higher chemical potential values. Notice
that from a determined chemical potential value, a multiple extrema structure takes place
in $\partial\Phi/\partial T$, different from the $\partial\rho_s/\partial T$ case, that
present only one maximum for any chemical potential, see Fig.~\ref{peaks}b. For those
$\partial\Phi/\partial T$ curves in which this effect is exhibited, the first peaks always
coincide with those in the $\partial\rho_s/\partial T$ curve at the same
$\mu_q$. The reason can be understood from the coupling between Eqs.~(\ref{mass})
and~(\ref{loop}). The peak in $\partial\Phi/\partial T$ coming from the abrupt fall in the
quark condensate, is reflected in the Polyakov loop via Eq. (\ref{loop}), since
$E=\sqrt{k^2+(m_0-2G_s\rho_s)^2}$. This variation in $\rho_s$ influences $\Phi$ also
generating an abrupt change in its value and, consequently, a peak in
$\partial\Phi/\partial T$. The second peak is uniquely coming from the Polyakov loop
dynamics itself since in this temperature range, the quark condensate practically
vanishes. This possibility of more than one peak in the temperature derivatives of the
order parameters was already reported for the PNJL model~\cite{tkahara}, as well as
in the linear sigma model coupled with the Polyakov loop~\cite{tkahara,marko,mao}.
\begin{figure}[!htb]
\centering
\includegraphics[scale=0.32]{peaks-a.eps}
\end{figure}
\vspace{-0.3cm}
\begin{figure}[!htb]
\centering
\includegraphics[scale=0.32]{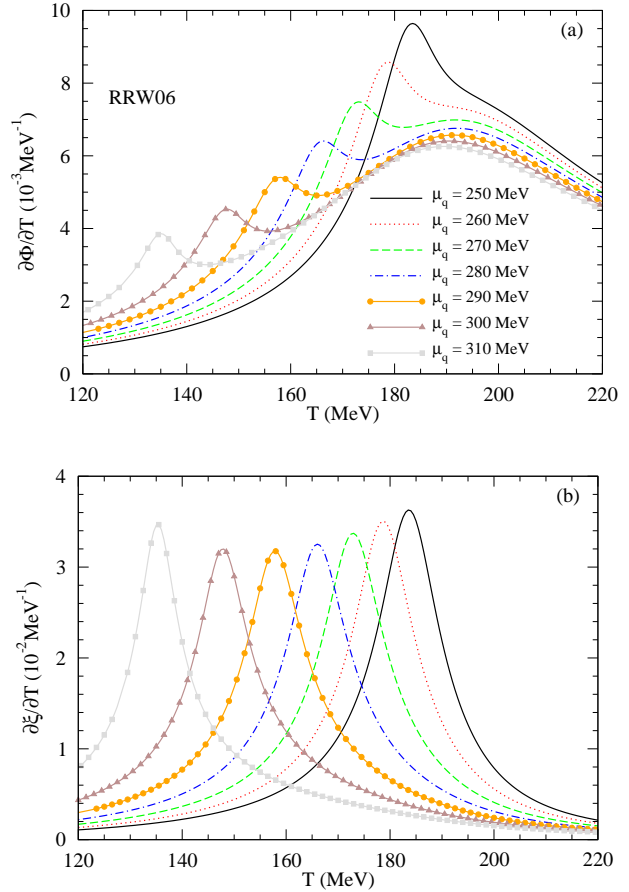}
\caption{(Color online) Temperature derivatives of a) $\Phi$ and b)
$\xi=\rho_s/\rho_{s(\mbox{\tiny vac})}$, as a function of $T$ at finite $\mu_q$ values.}
\label{peaks}
\end{figure}

We point out here that the construction of the phase diagram based on the choice of the
coincident peaks in $\partial\rho_s/\partial T$ and $\partial\Phi/\partial T$, leads to a
situation where the region in which the chiral symmetry is broken (restored) and that one
in which the quarks are confined (deconfined), are exactly the same. Therefore, there is
no possibility to identify, following this criterium, a quarkyonic phase~\cite{mclerran},
region where the chiral symmetry is restored but with quarks still confined.

Another criterium used to construct the PNJL diagrams, is investigate the
magnitude of the order parameters, since their values are directly related to
the symmetries that are broken or not in the regions delimited by the
boundary curves in the $T\times\mu_q$ plane. It is well know that $\rho_s\neq
0$ indicates broken chiral symmetry, and $\rho_s=0$ means that this
symmetry is restored. The same concept is adopted for the Polyakov loop $\Phi$.
The difference is that the involved symmetry is the center symmetry, closely
associated with the confinement phenomenon~\cite{weiss}. In this case, the
pseudo-critical temperature is defined by Fukushima~\cite{fuku2} as that in which all the
order parameters, normalized by its vacuum values (with exception for the
Polyakov loop), reach the value of $1/2$. Thus, the author constructed three
distinct boundary curves, also identifying the quarkyonic phase, in this case
for the SU(3) version of the PNJL model, see Fig. 12 of Ref.~\cite{fuku2}.
\onecolumngrid
\textcolor{white}{asdfas}

Our purpose here is furnish an alternative and more natural criterium to identify the
different quark phases and its boundary curves, but also using the magnitude of $\rho_s$
and $\Phi$. The method is based on the analysis of the projected surface of the order
parameters as a function
of $T$ and $\mu_q$. An example is given in Fig.~\ref{log3d} for the RRW06
parametrization.
\vspace{-0.8cm}
\begin{figure}[!htb]
\centering
\includegraphics[scale=0.65]{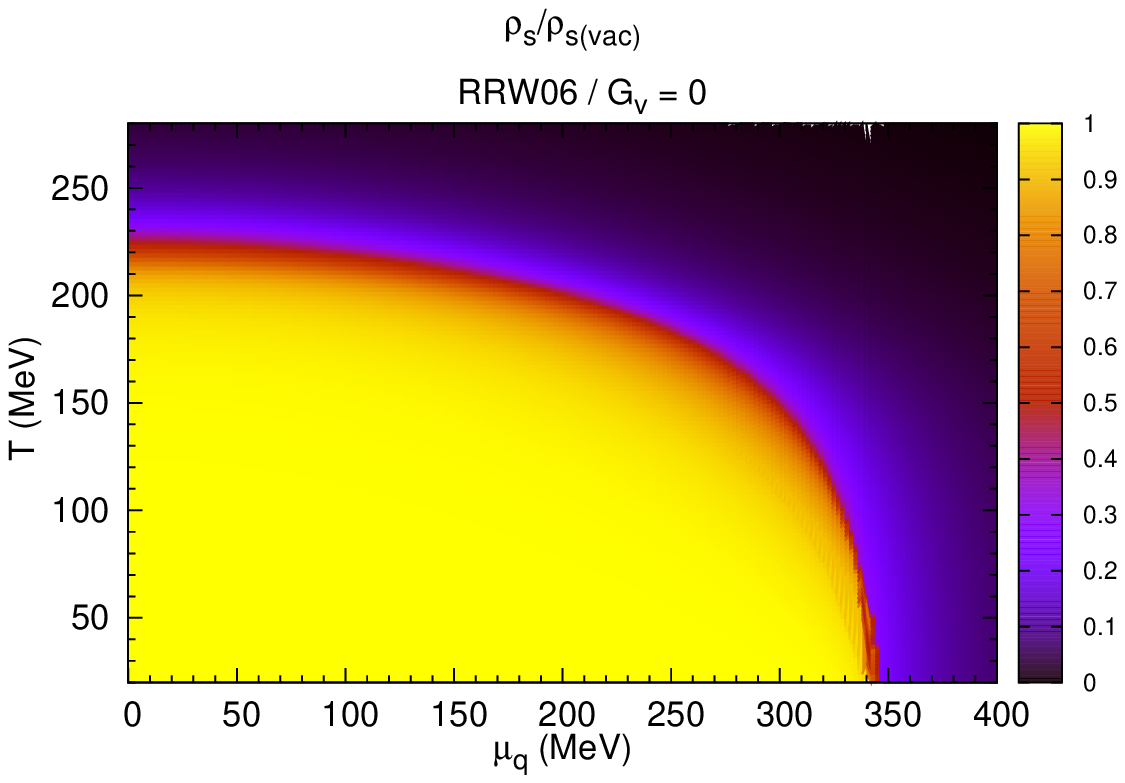}
\includegraphics[scale=0.65]{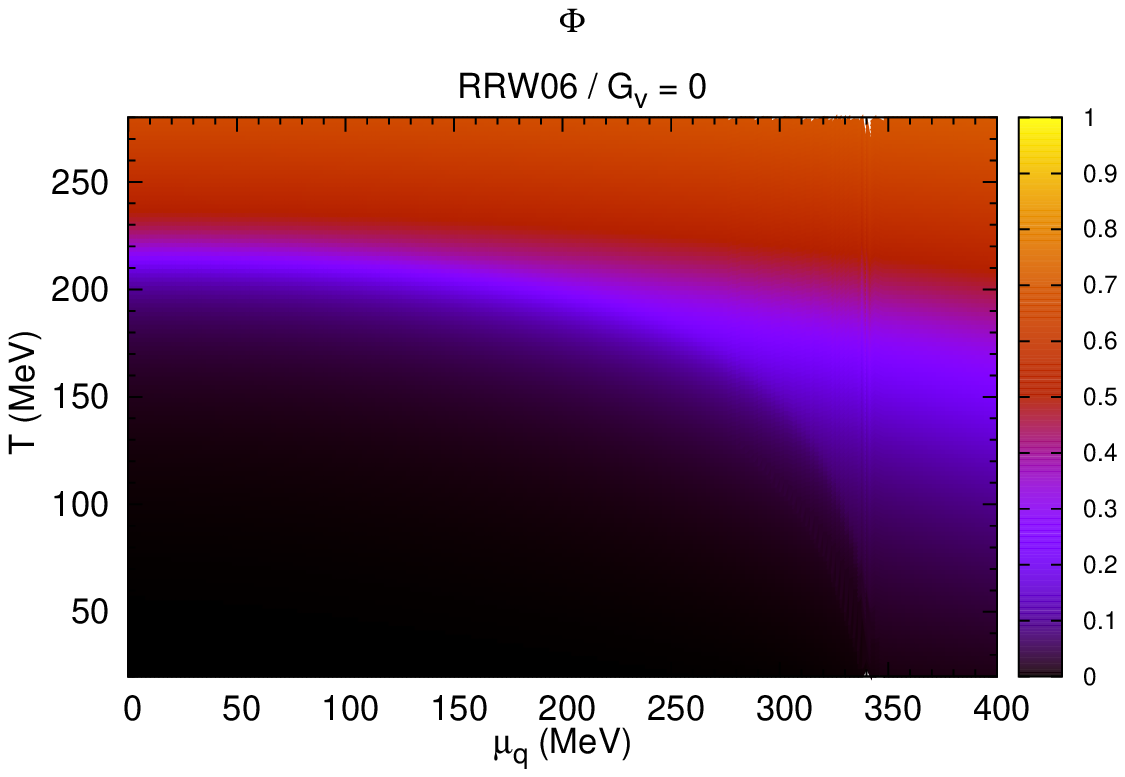}
\vspace{-0.5cm}
\caption{(Color online) Order parameters surfaces $\rho_s$ (left panel) and $\Phi$ (right
panel) projected in the $T\times\mu_q$ plane.}
\label{log3d}
\end{figure}

The broken and restored chiral symmetry regions are very well defined, as well as its
boundary in the left panel. In the $\Phi$ plot (right panel), one can also recognize the
confined and deconfined quark phases. The interesting feature in this diagram is the
natural emergence of a phase between the confined and the deconfined one. To become clear
that such a phase is the quakyonic one, we plot inside these diagrams, the curves
delimiting all regions. The result is shown in Fig.~\ref{log3d-all}.
\vspace{-0.5cm}
\begin{figure}[!htb]
\centering
\includegraphics[scale=0.67]{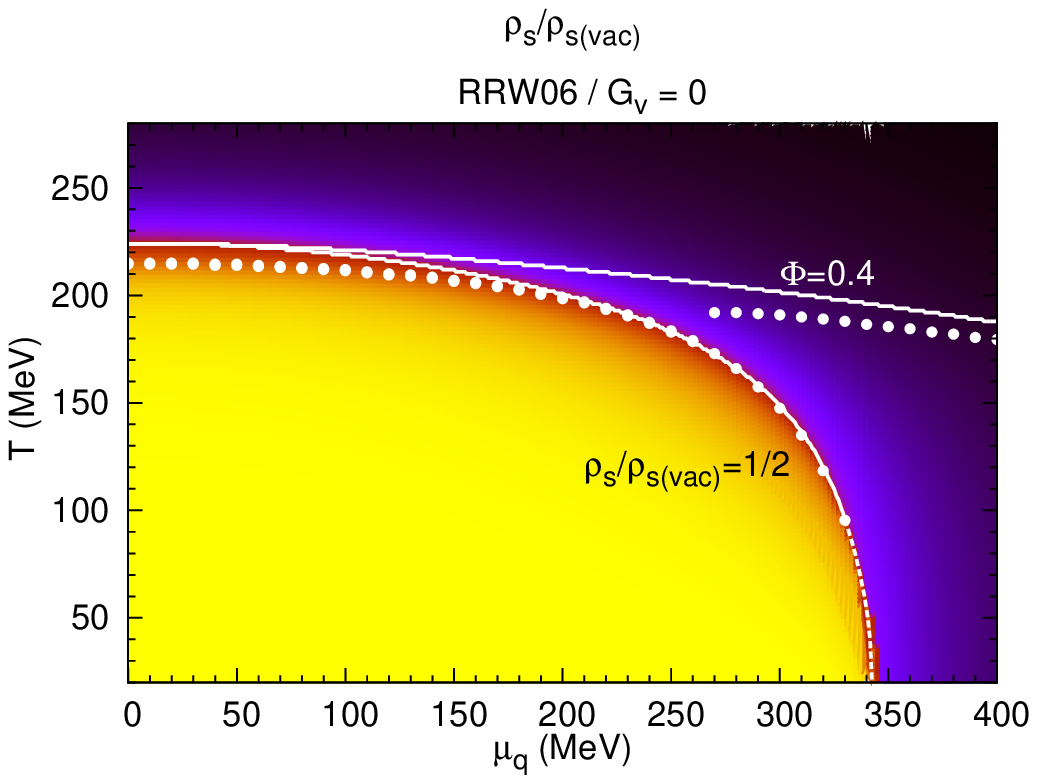}
\includegraphics[scale=0.67]{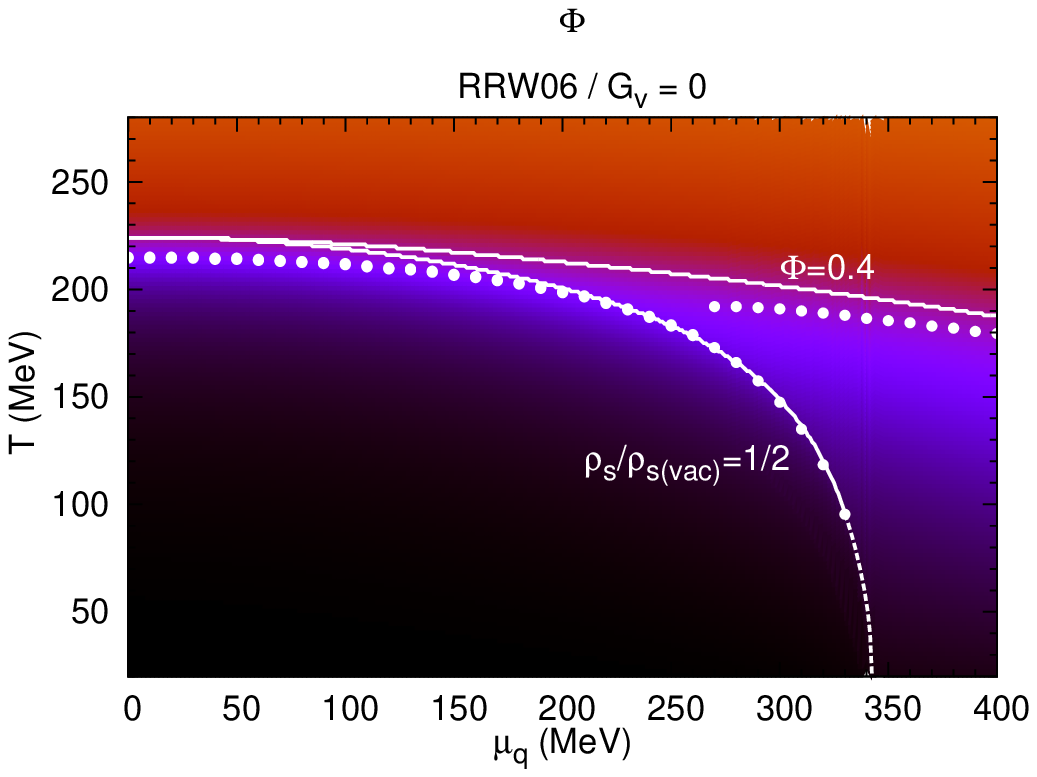}
\vspace{-0.5cm}
\caption{(Color online) The same as in Fig.~\ref{log3d}. Full lines: curves
corresponding to the points in which $\rho_s/\rho_{s(\mbox{\tiny vac})}=1/2$ and
$\Phi=0.4$. Full circle curves: obtained from the peaks in $\partial\rho_s/\partial T$ and
$\partial\Phi/\partial T$. Dashed lines: first order phase transitions curves.}
\label{log3d-all}
\end{figure}
\twocolumngrid

The lower full curves were generated by making fixed the value of
$\rho_s/\rho_{s(\mbox{\tiny vac})}$. From Fig.~\ref{log3d}, the color code suggests that
the $1/2$ value is a good representative of the boundary between the broken and restored
chiral symmetry phases. In the case of the Polyakov loop curves, upper full ones, we
found $\Phi=0.4$ as a reasonable value to delimit the onset of the deconfined quark phase.
We remark here that other values could represent this boundary line, depending on the used
model. The authors of~\cite{scoccola} (see their Figs. 1 and 5) used a value of
$\Phi=0.3$ for the nonlocal version of the PNJL model in the chiral limit, and for finite
quark masses, but using a Landau expansion and susceptibilities to find the crossover
chiral line. Also Fukushima defined $\rho_s/\rho_{s(\mbox{\tiny vac})}=\Phi=1/2$
to represent the $\rho_s/\rho_{s(\mbox{\tiny vac})}$ and $\Phi$ boundaries in the
PNJL-SU(3) model~\cite{fuku2}, claiming that the magnitude of the order parameter is a
more suitable quantity to probe the physical state of matter. Here, we construct the
boundary lines by finding the suitable values of the order parameters from their
projection on the $T\times\mu_q$ plane.

From Fig.~\ref{log3d-all} is clear that the region between the two full
lines is the quarkyonic phase, exhibited in our calculations even in the SU(2)
version of the PNJL parametrizations presented here. In the same figure we also
furnishes the first order transition lines (dashed ones).
In order to clarify the definition of the solid boundaries in Fig. \ref{log3d-all},
we present in Fig. \ref{muqeq0} the temperature dependence of the order parameters for
$\mu_q=0$. In this figure, we show by the circles the peaks positions of
$\partial\Phi/\partial T$ and $\partial\rho_s/\partial T$. The position of the points in
which $\Phi=0.4$ and $\rho_s/\rho_{s(\mbox{\tiny vac})}=1/2$ are denoted by the crosses.
\begin{figure}[!htb]
\centering
\includegraphics[scale=0.31]{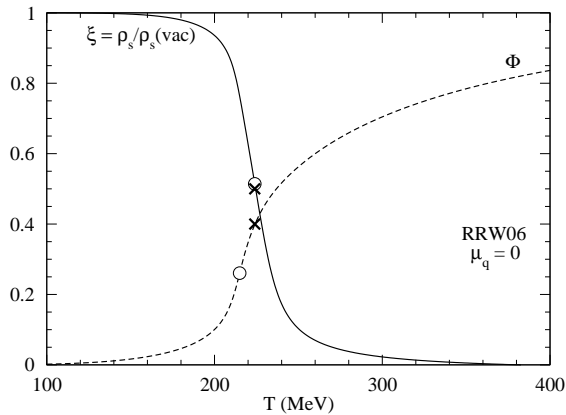}
\caption{Order parameters as a function of $T$. Circles: peaks positions of
$\partial\Phi/\partial T$ and $\partial\rho_s/\partial T$. Crosses: points positions of
$\Phi=0.4$ and $\rho_s/\rho_{s(\mbox{\tiny vac})}=1/2$.}
\label{muqeq0}
\end{figure}

In order to compare this new method of construction of the boundary lines with those that
uses the peaks criterium, we also display in Fig.~\ref{log3d-all} the points corresponding
to the peaks of $\partial\rho_s/\partial T$ and $\partial\Phi/\partial T$. In the range of
$0<\mu_q\lesssim 270$~MeV, the derivative curves present only one peak that are almost
coincident, see Fig.~\ref{peaks}. Therefore, it is possible to determine only one curve in
the $T\times\mu_q$ plane (circles starting at $\mu_q=0$ in Fig.~\ref{log3d-all}). From
$\mu_q\sim 270$~MeV, the $\partial\Phi/\partial T$ curve presents two peaks, one of
them, the first one, the same as in the $\partial\rho_s/\partial T$ curve. The second
peaks of $\partial\Phi/\partial T$ are represented in Fig.~\ref{log3d-all} by the circles
starting at $\mu_q\sim 270$~MeV. The composition of the first and second peaks lines
leads to delimit a smaller quarkyonic phase when we compare it to that obtained by the
region between the two full lines. This feature is also verified when we add an additional
vector interaction in the PNJL model.

In the cases in which one chooses only the first peaks in the $\partial\Phi/\partial T$
curve, the circles starting at $\mu_q\sim 270$~MeV would not appear and consequently,
there would be no quarkyonic phase. In this case, the quark phase diagram would lose an
essential information. This does not happen if we use the method of the magnitude of the
order parameters to construct the boundary curves. The quarkyonic phase is always present
in the phase diagram.

We have checked that using $\Phi\neq\Phi^*$ the phase diagrams are not altered
significantly. We illustrate our computation in Fig.~\ref{phiphistar} for the potential
RRW06, which should be compared with Fig.~\ref{log3d}. We present results of the phase
diagrams for $\rho_s$, $\Phi$ and $\Phi^*$. We observe negligible differences in the
$\Phi$ and $\Phi^*$ projections, compared to the $\Phi$ one for the $\Phi=\Phi^*$
case showed in Fig.~\ref{log3d}. The same pattern is verified in the case of $\rho_s$. 
\vspace{-0.6cm}
\begin{figure}[!htb]
\centering
\includegraphics[scale=0.65]{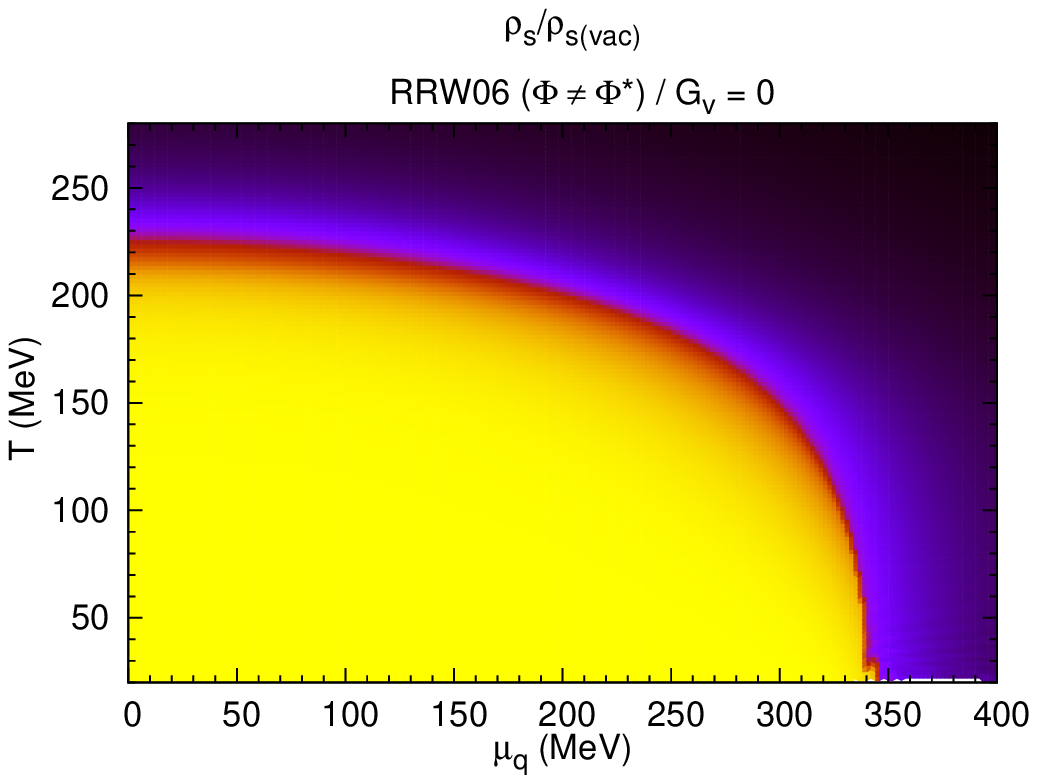}\\
\vspace{-0.5cm}
\includegraphics[scale=0.65]{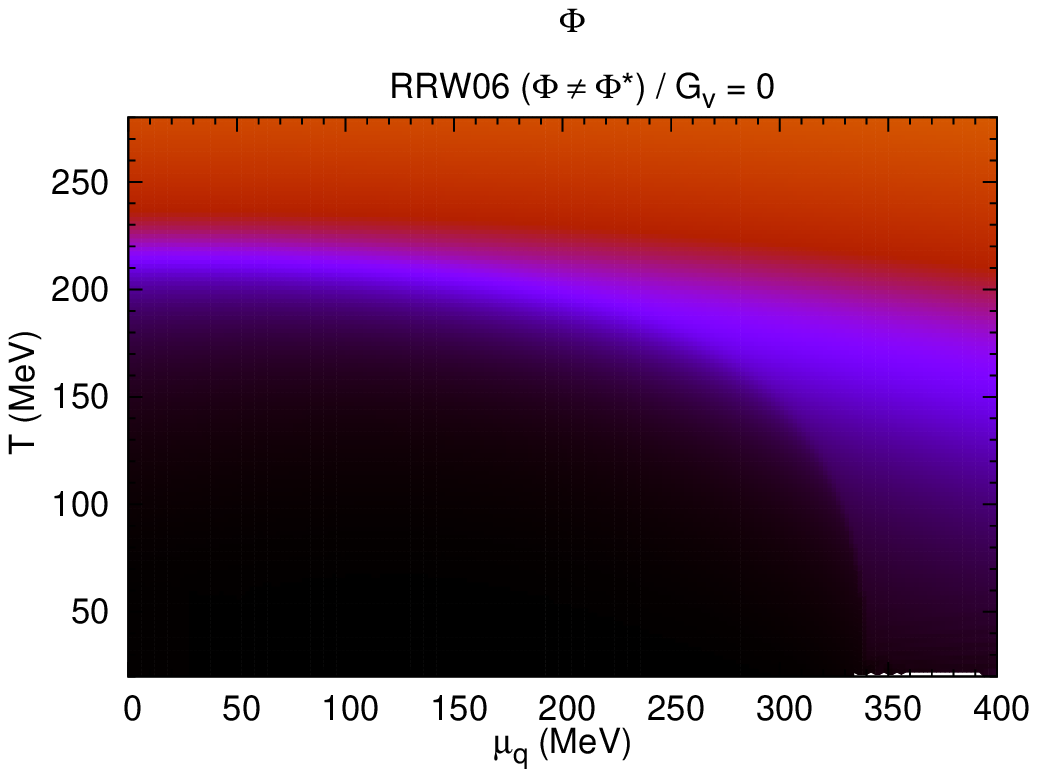} \\
\vspace{-0.5cm}
\includegraphics[scale=0.65]{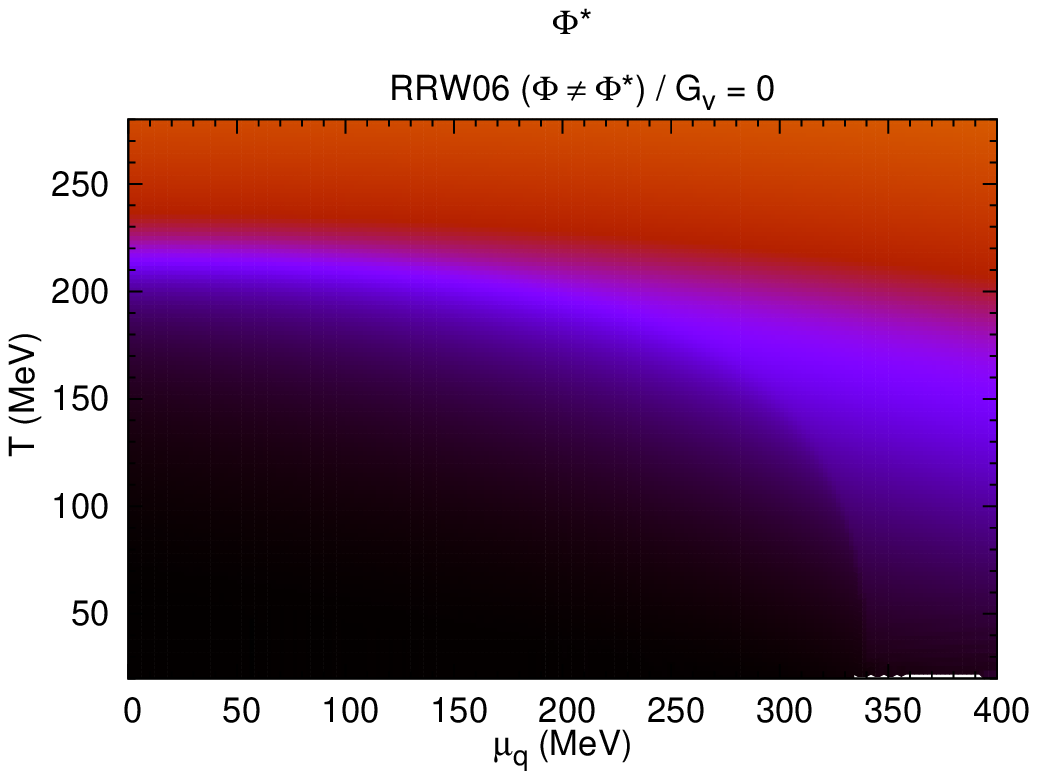}
\vspace{-0.5cm}
\caption{(Color online) Order parameters surfaces for $\rho_s$ (upper panel), $\Phi$ (middle
panel)  and $\Phi^*$ (lower
panel) projected in the $T\times\mu_q$ plane.}
\label{phiphistar}
\end{figure}

Our results corroborate the findings of Ref. \cite{weise1} for a single $\mu_q$
value, giving  support to the assumption  $\Phi=\Phi^*$ used in our work for a large
region of $\mu_q>0$, necessary for the study of PNJL phase diagrams.

\subsection{Effect of the vector interaction}
\label{discussion-vec}

It is known that a vector-type interaction in the PNJL model, is responsible
to shrink the first order phase transition~\cite{fuku2,kbt,fuku3}.
Therefore, the critical end point is moved in the direction to be completely
removed, as the strength of the interaction is increasing. The same effect could
also be observed in the NJL model~\cite{buballa,kashiwa2}. The inclusion of a
vector term of the form $\,-G_V(\bar{q}\gamma^\mu q)^2\,$ in the PNJL Lagrangian density
modifies the grand canonical potential as
\begin{eqnarray}
\Omega_{\mbox{\tiny PNJL}}(\mu_q,T,\Phi)\rightarrow \Omega_{\mbox{\tiny
PNJL}}(\tilde{\mu_q},T,\Phi)-G_V\rho^2,
\end{eqnarray}
with
\begin{eqnarray}
\tilde{\mu_q} = \mu_q - 2G_V\rho,
\label{effmu}
\end{eqnarray}
being the effective chemical potential, and $\,\rho\,$ the quark
density. All the other quantities and equations are modified by making
$\,\mu_q\rightarrow\tilde{\mu_q}\,$. Therefore, besides the Eqs. (\ref{mass}) and
(\ref{loop}), also the Eq. (\ref{effmu}) should be take into account in the
self-consistent solutions of the order parameters.
\begin{figure}[!htb]
\centering
\includegraphics[scale=0.32]{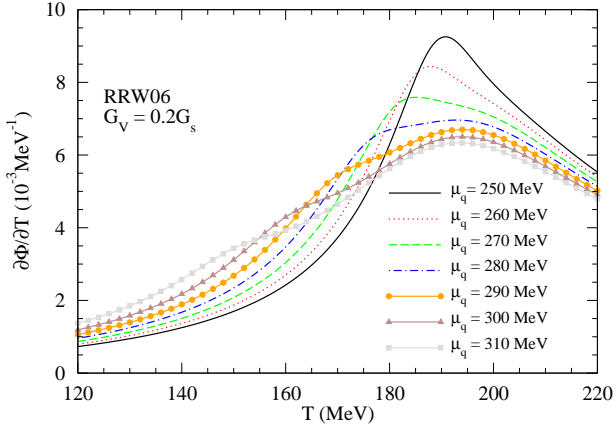}
\caption{(Color online) Temperature derivative of the order parameter $\Phi$ as a
function of the temperature for $G_V=0.2G_s$.}
\label{loggv02}
\end{figure}
\vspace{-0.8cm}
\begin{figure}[!htb]
\centering
\includegraphics[scale=0.68]{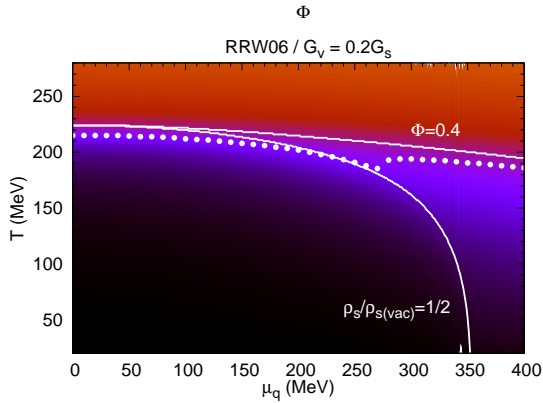}
\vspace{-1cm}
\caption{(Color online) $\Phi$ projection at the $T\times\mu_q$ plane for $G_V=0.2G_s$.}
\label{log3d-gv01}
\end{figure}

Other effect of the vector interaction is to change the double peak structure
in $\partial\Phi/\partial T$, compared to the case in which $G_V=0$. Fig.~\ref{loggv02} shows
this behavior for the $G_V=0.2G_s$ case. Notice that in this case there is only one peak. The
boundary curve constructed from the analysis of $\partial\Phi/\partial T$ is shown in the
$\Phi$ projected curve in the $T\times\mu_q$ plane of Fig.~\ref{log3d-gv01}. In the same
figure, we also show the boundary lines constructed by taking the fixed values of
$\rho_s/\rho_{s(\mbox{\tiny vac})}=1/2$ and $\Phi=0.4$.

It is clear that if the peaks criterium in $\partial\Phi/\partial T$ is adopted in this
situation, the obtained curve is not sufficient to correctly delimit all the possible
phases of the system, as in the case of $G_V=0$, see Fig.~\ref{log3d-all}. Therefore, it
is also necessary to use the peaks of $\partial\rho_s/\partial T$ to make clear the
distinct regions. Notice also the difference between the curve obtained via
$\partial\Phi/\partial T$ peaks and that constructed via $\Phi=0.4$. The latter one is
more precise in the description of the boundary of the confinement/deconfinement phases.
\vspace{-0.3cm}
\begin{figure}[!htb]
\centering
\includegraphics[scale=0.35]{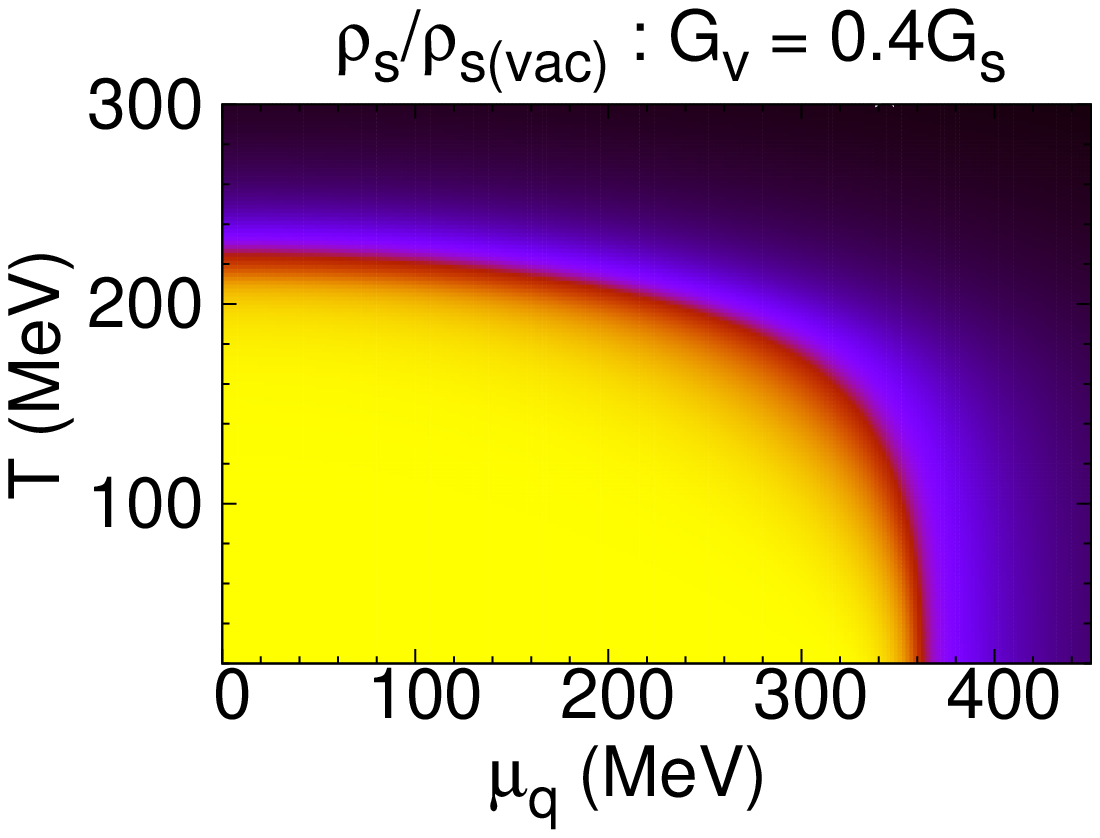}
 \hspace{-0.5cm}
\includegraphics[scale=0.35]{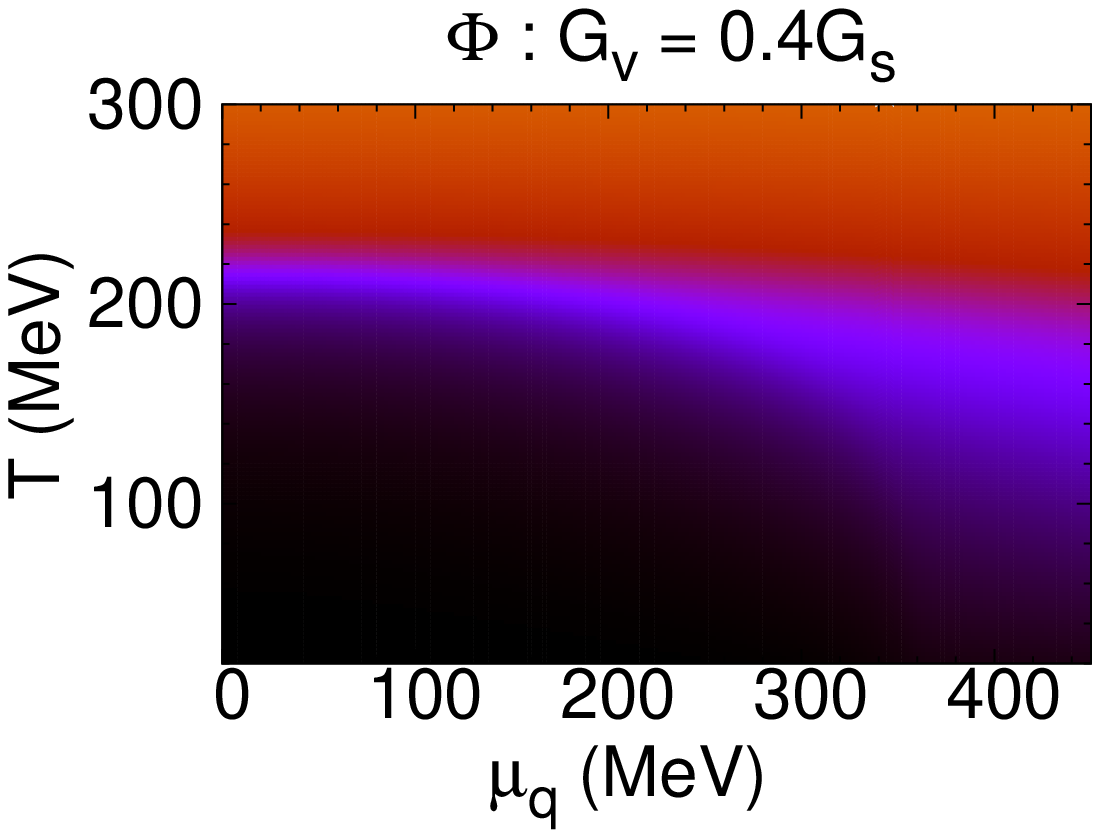}
\includegraphics[scale=0.35]{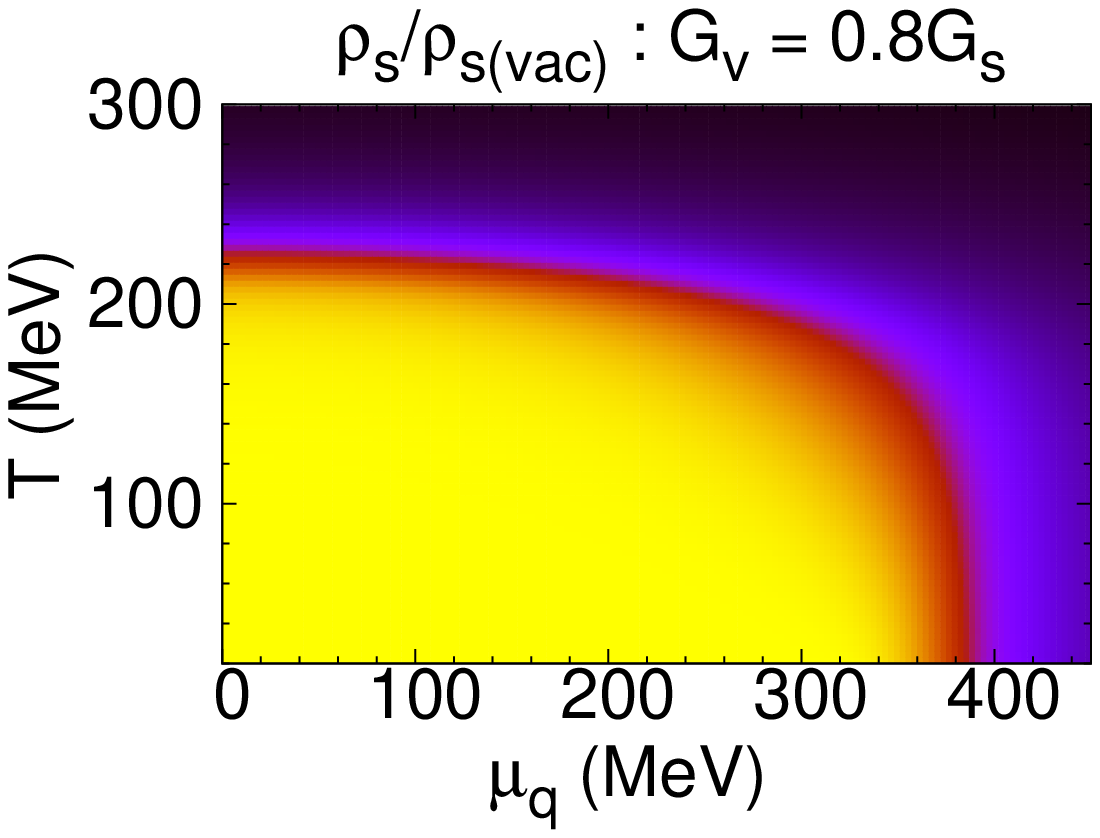}
 \hspace{-0.5cm}
\includegraphics[scale=0.35]{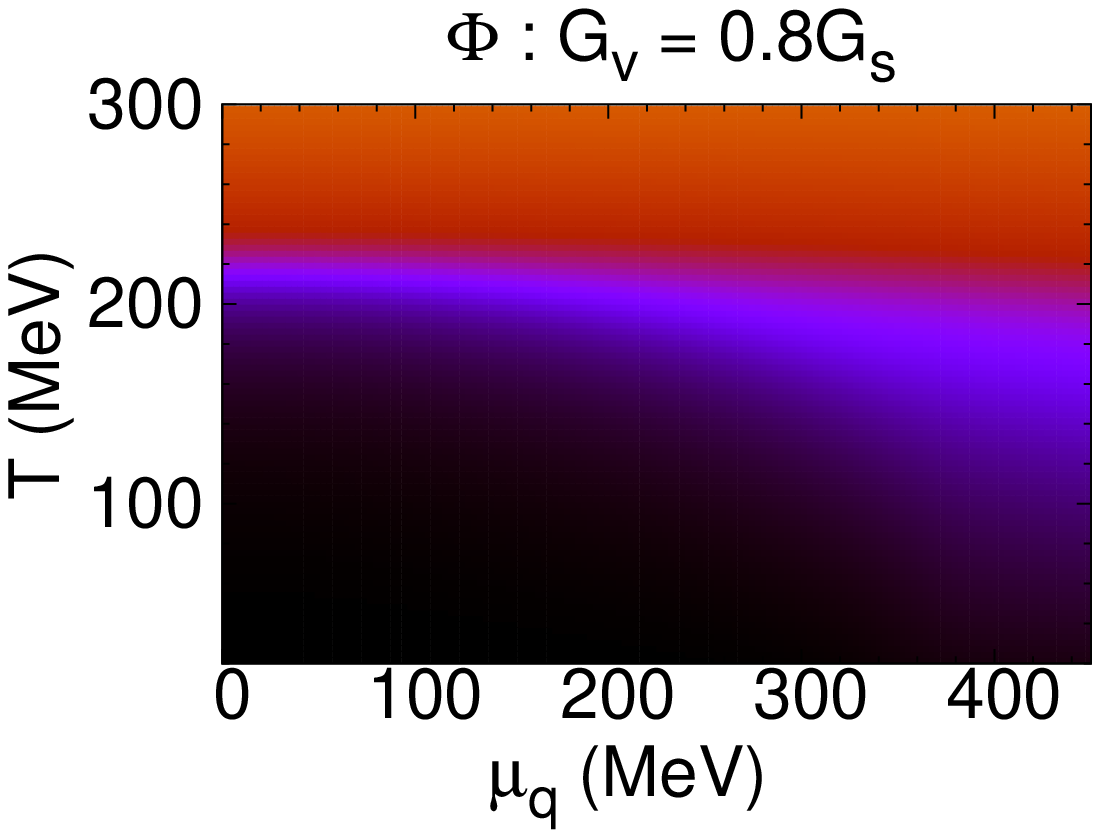}
\includegraphics[scale=0.35]{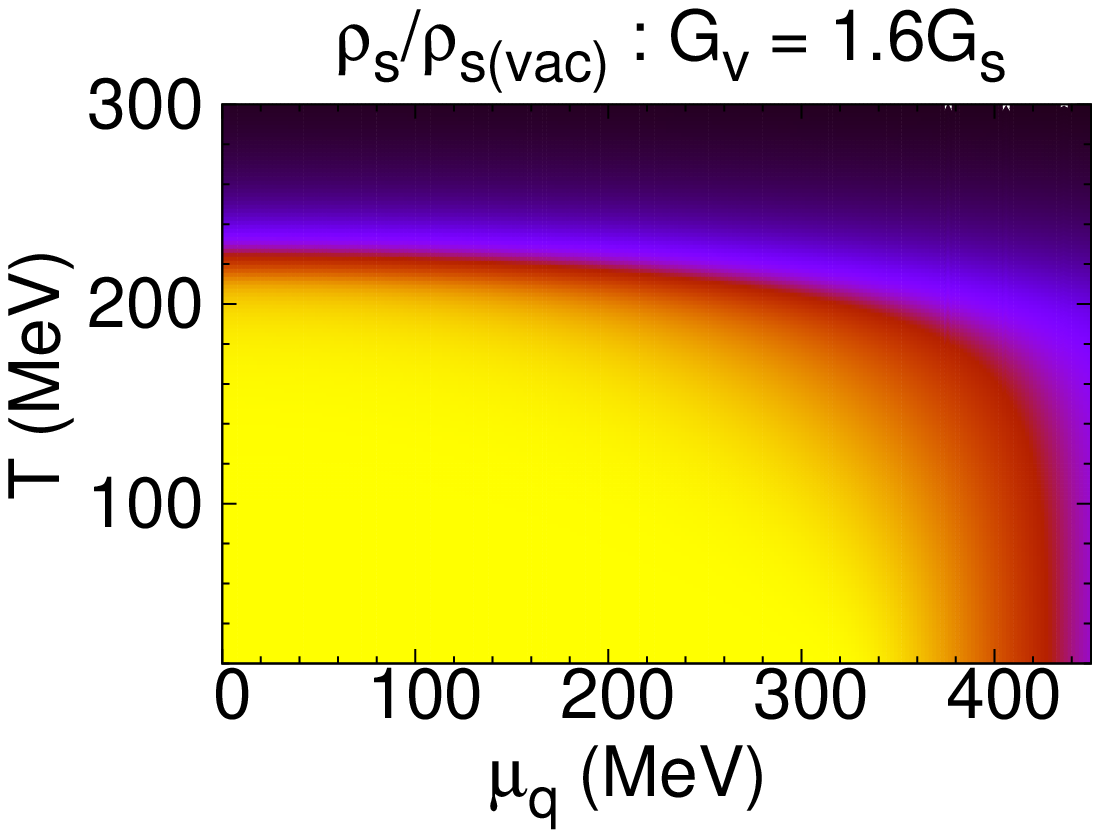}
 \hspace{-0.5cm}
\includegraphics[scale=0.35]{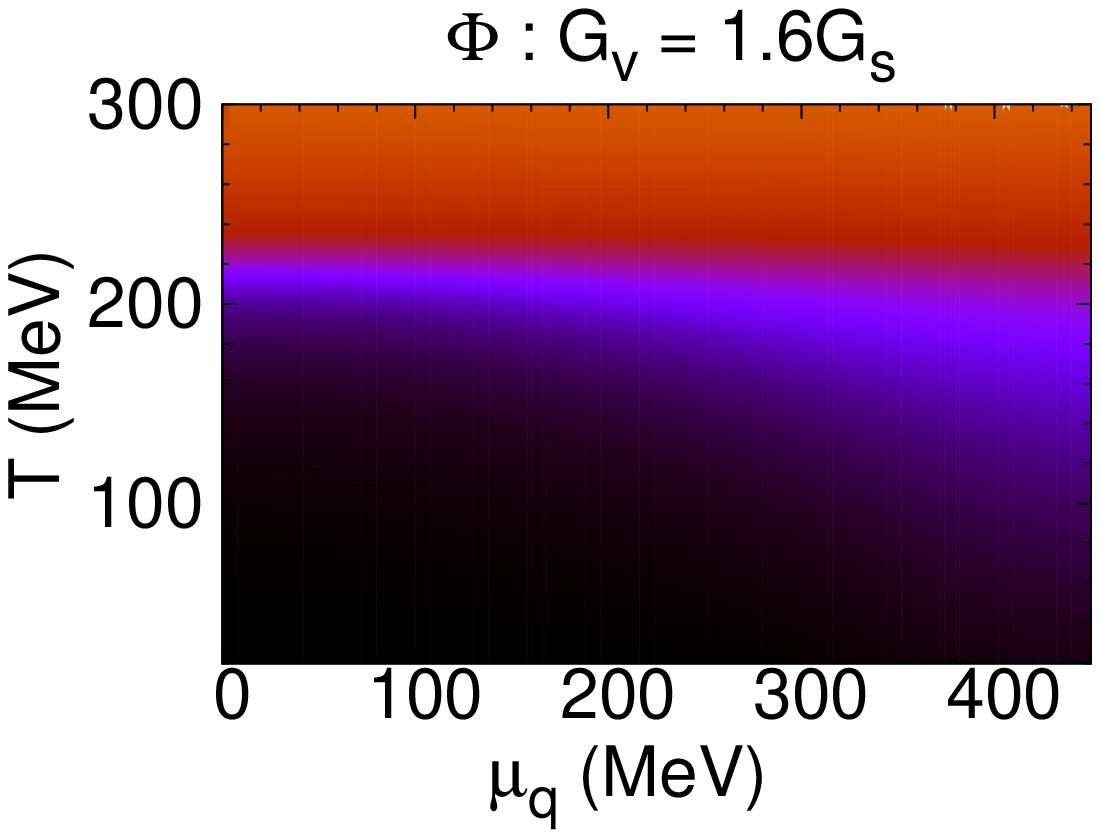}
\vspace{-0.1cm}
\caption{(Color online) Projection of the order parameter surfaces $\rho_s$ (left panels)
an $\Phi$ (right panels) in the $T\times\mu_q$ plane for different $G_V$ values in the
RRW06 potential.}
\label{gvevolution}
\end{figure}

Finally, we show how the strength of the vector interaction affects the quarkyonic phase
in the PNJL model. Fig.~\ref{gvevolution} shows the behavior of the order parameters
for some values of $G_V$. The pattern exhibited shows that the quarkyonic phase tends to
become smaller, as $G_V$ is increased.

Even with such bands, showed when $G_V$ is increased, it is still possible to use the
magnitude of the order parameters to define boundaries of the broken/restored chiral
symmetry, and confinement/deconfinement phases. Indeed, we have studied in
Ref.~\cite{prd2} the quark phase diagrams of PNJL models constructing the boundary of
broken/restored chiral symmetry phase, for distinct $G_V$ values, using different values
of $\rho_s/\rho_{s(\mbox{\tiny vac})}$. Notice also a diffusing effect on in the values of
$\rho_s$, mainly at higher values of $\mu_q$. A similar behavior is also verified for
different values of the $T_0$ parameter in the PNJL models even at $G_V=0$. In next
subsection, we show the case for $T_0=205$~MeV in the RRW06 parametrization.

Notice also that the effect of moving the boundary related to broken/restored chiral
symmetry phases is not observed in the $\Phi$ projection. In this case, the boundary of
confined/deconfined phases remains unchanged. The quarkyonic phase is moved to the
direction of increasing $\mu_q$ values, but the phase related to free massless quarks is
unaffected.

As a last remark of the inclusion of the vector interaction, we point out that the
projections of the order parameters generated by the saddle point approach and by the method
of Ref.~\cite{mintz} are exactly the same, since for the values of $G_V$ used here, there is
no regions of first order phase transitions and therefore, no regions of negative eigenvalues
of the Hessian matrix.

\subsection{Effect of the $T_0$ parameter and the EPNJL model}

The Polyakov potentials presented in Eqs. (\ref{rtw05})-(\ref{fuku08}) have their free
parameters adjusted to correctly reproduce some lattice QCD results for the pure gluon
sector (quenched approximation). In particular, the value of $T_0=270$~MeV is the
temperature in which the gluonic system presents a first order phase transition. The
discontinuity in this case is verified in the Polyakov loop plotted as a function of
$T$. With the found parameters, the PNJL model is used to describe, in an effective way,
the system with quarks and gluons. However, the transition temperature found in PNJL
models at $\mu_q=0$ is higher than that obtained by lattice QCD calculations. The latter
is given by $173\pm 8$~MeV~\cite{latresults}. The former is calculated as
$T(\mu_q=0)>200$~MeV through the peaks criterium or even using the magnitude of the
order parameters (projection in the $T\times\mu_q$), see the starting point at $\mu_q=0$
of the circles and full curves in Fig.~\ref{log3d-all}.

In order to make the PNJL model consistent with the lattice results at $\mu_q=0$, the
rescaling in $T_0$ is often used in the literature. The change to $T_0=190$~MeV decreases
the transition temperature of PNJL models, at zero chemical potential, to compatible
values when the peaks criterium is adopted. However, as pointed out in
Ref.~\cite{weise1}, the peaks of $\partial\rho_s/\partial T$ and $\partial\Phi/\partial T$
are not coincident anymore as in the case in which $T_0=270$~MeV. Due to lattice QCD
studies indicate that quark deconfinement and chiral restoration occurs at same
temperature at $\mu_q=0$, this problem is circumvented in PNJL model, by taking the
average temperatures associated to the peaks of $\partial\rho_s/\partial T$ and
$\partial\Phi/\partial T$. Such a procedure generates only one boundary curve in the
$T\times\mu_q$ plane, characterizing one region in which: i) chiral symmetry is broken
with confined quarks, and other one presenting: ii) restoration of chiral symmetry and
deconfined quarks. There is no possibility of quarkyonic phase (restored chiral symmetry
and confined quarks). This is not the case if we construct the boundary curves from the
analysis of the $\rho_s$ and $\Phi$ projections, as we will make clear below.
\vspace{-0.4cm}
\begin{figure}[!htb]
\centering
\includegraphics[scale=0.65]{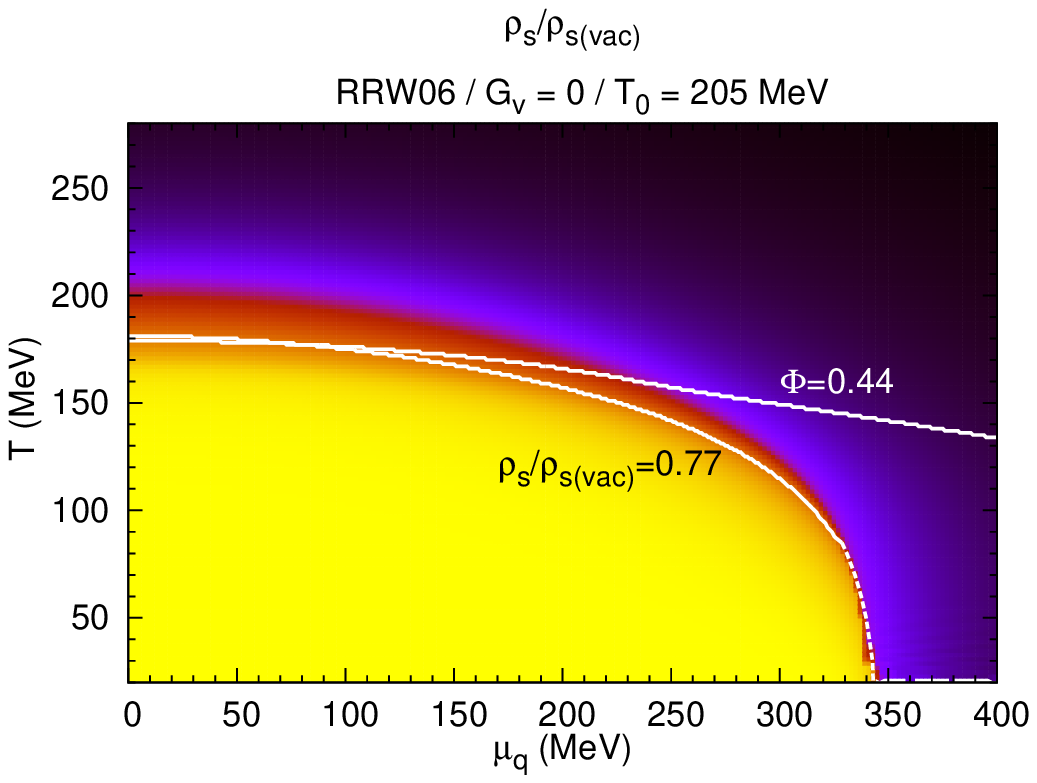}
\end{figure}
\vspace{-1.5cm}
\begin{figure}[!htb]
\includegraphics[scale=0.65]{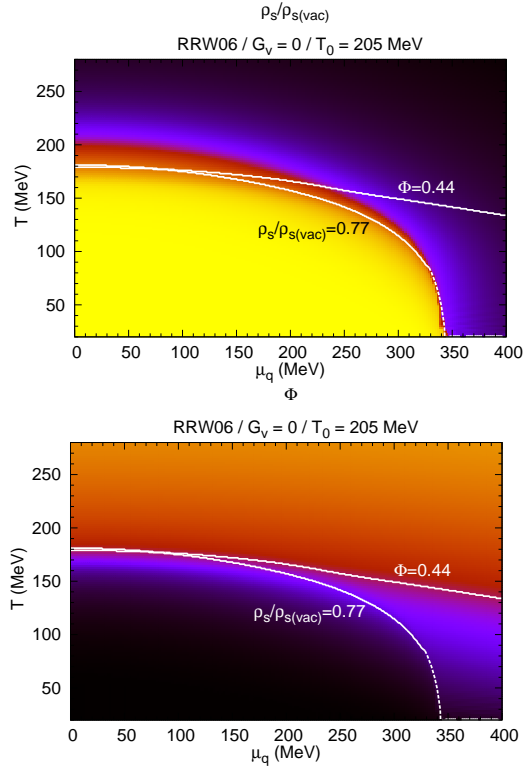}
\vspace{-0.5cm}
\caption{(Color online) $\rho_s$ (upper panel) and $\Phi$ (lower panel) surfaces projected
in the $T\times\mu_q$ plane
for RRW06 parametrization with $T_0=205$~MeV.}
\label{log3dt0205}
\end{figure}

If the projection of the order parameters in the $T\times\mu_q$ plane is used, it is
also possible find suitable values of the transition temperature in PNJL model at
$\mu_q=0$. This is done in Fig.~\ref{log3dt0205} by rescaling $T_0$ from $270$~MeV to
$205$~MeV.

Important points deserve to be discussed regarding the phase diagrams exhibited in this
figure. First of all, notice that the change in the $T_0$ parameter makes the boundary of
the broken/restored chiral symmetry phases larger than that presented in Fig.~\ref{log3d}.
The ``red line'' in Fig.~\ref{log3d} gives rise to the ``red band'' in
Fig.~\ref{log3dt0205}. It means that if we want to construct a boundary curve in the
$T\times\mu_q$ plane, it has to be inside such a band. The same does not occur for the
$\Phi$ projection. Notice that we still can define unambiguously, a curve separating the
confined and deconfined phases. Indeed, such a curve is constructed by making
$\Phi=0.44$, value that furnishes a curve in which the transition temperature at $\mu_q=0$
is compatible to lattice QCD results. Since this curve is defined, one has to find a fixed
value of $\rho_s/\rho_{s(\mbox{\tiny vac})}$ in order to make the boundary curve of the
broken/restored chiral symmetry phases present the same transition temperature at
$\mu_q=0$. The value found in this case is $\rho_s/\rho_{s(\mbox{\tiny vac})}=0.77$.

The choice of $\rho_s/\rho_{s(\mbox{\tiny vac})}=0.77$ is not unique. Different values
that make the boundary curve inside the red band can be found. However, these values
generate curves that do not have the same transition temperature at $\mu_q=0$ as that
presented by the $\Phi=0.44$ curve, and also temperatures different from
$T(\mu_q=0)\sim170$~MeV. Therefore, not satisfying the constraint established by the
lattice results.
\vspace{-0.5cm}
\begin{figure}[!htb]
\centering
\includegraphics[scale=0.65]{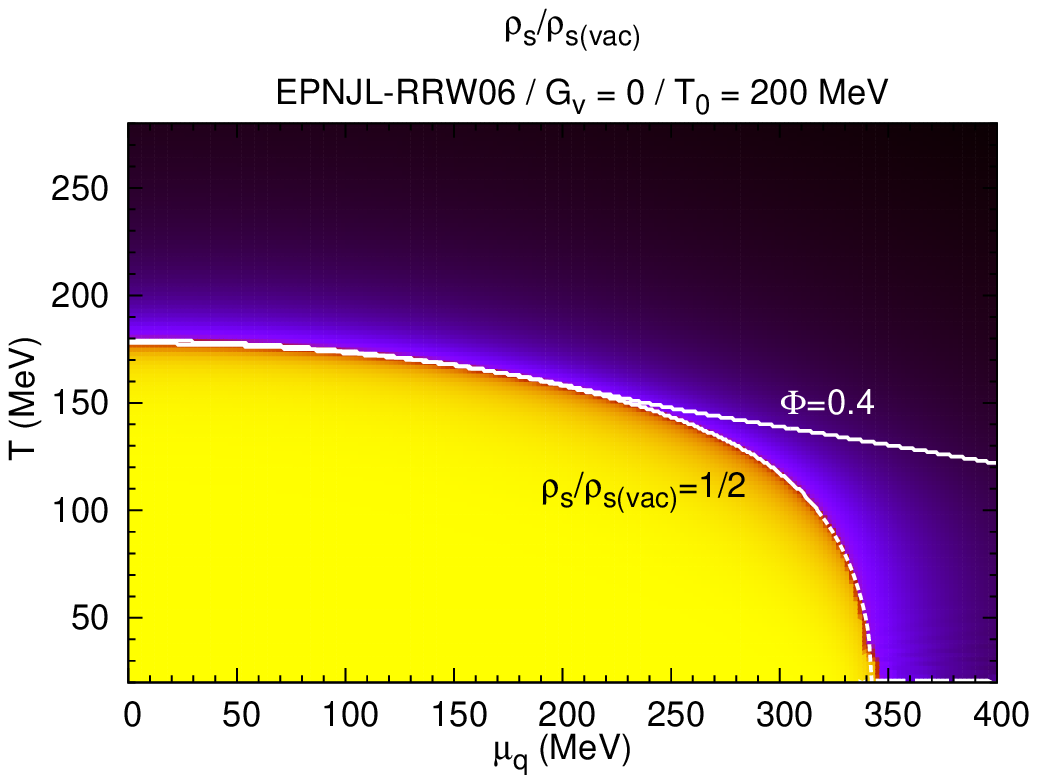}
\end{figure}
\vspace{-1.4cm}
\begin{figure}[!htb]
\includegraphics[scale=0.65]{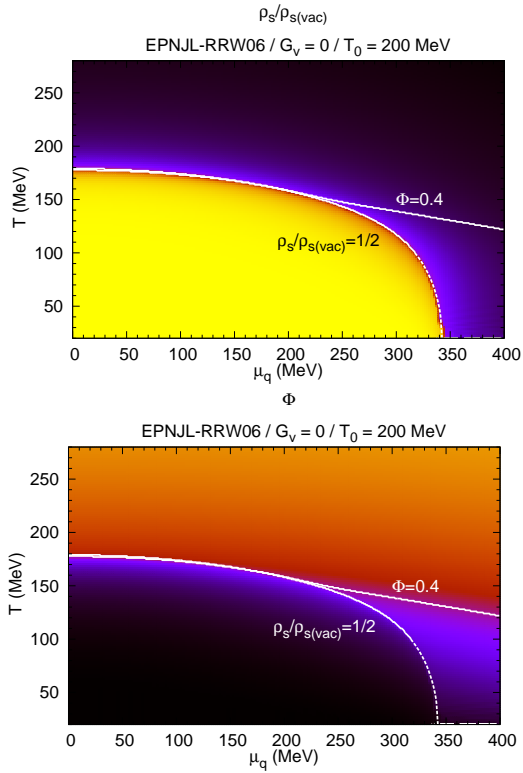}
\vspace{-0.5cm}
\caption{(Color online) $\rho_s$ (upper panel) and $\Phi$ (lower panel) surfaces of the
EPNJL model, projected in the $T\times\mu_q$ plane for the RRW06 parametrization with
$T_0=200$~MeV.}
\label{log3dt0200}
\end{figure}

An important aspect showed in Fig.~\ref{log3dt0205} is that the change in the $T_0$
parameter makes the confined phase smaller than that presented in the case in which $T_0$
has its original value of $270$~MeV. Actually, this is the reason of the transition
temperature at $\mu_q=0$ be compatible with lattice results. The same is not verified for
the broken chiral symmetry phase. The change in the values of $\Phi$, induced by the $T_0$
rescaling, is not totally followed by $\rho_s$. As already mentioned, the boundary of the
phases in the $\rho_s$ projection on the $T\times\mu_q$ plane becomes a band that is still
larger for lower values of $T_0$. This is due to the ``weak interaction'' between the
order parameters $\Phi$ and $\rho_s$ presented in the structure of the PNJL models treated
here. This shortcoming in PNJL models can be circumvented by including a $\Phi$
dependence in the scalar coupling $G_s$, i. e., by making $G_s=G_s(\Phi)$. In order
to illustrate the effect of such a modification, we use the $\Phi$ dependence on $G_s$
as given by,
\begin{eqnarray}
G_s(\Phi) = G_s[1-\alpha_1\Phi\Phi^* - \alpha_2(\Phi^3 + \Phi^{*3})],
\end{eqnarray}
closely following Ref.~\cite{epnjl1}, even concerning the values of
$\alpha_1=\alpha_2=0.2$. The PNJL model modified by making $G_s\to G_s(\Phi)$ is named
EPNJL model~\cite{epnjl2}, since $G_s=G_s(\Phi)$ is an effective vertex called
entanglement vertex~\cite{epnjl1}. In Fig.~\ref{log3dt0200}, we show the order parameters
projections of the RRW06 parametrization of the EPNJL model for $T_0=200$~MeV.

Notice that the consequence of the strong correlation between $\rho_s$ and $\Phi$ in
the phase diagrams of EPNJL model, is to reduce the red band to a line in the $\rho_s$
projection. In the case, the boundary line of the broken/restored chiral symmetry phases
is unambiguously given by $\rho_s/\rho_{s(\mbox{\tiny vac})}=1/2$, exactly as in the case
of the PNJL model for $T_0=270$~MeV, see Fig.~\ref{log3d-all}. Also, the boundary curve of
the confined/deconfined is defined by fixing the value of $\Phi=0.4$. Now, both curves
start at a transition temperature at $\mu_q=0$ comparable to the lattice result of
$173\pm8$~MeV.

Other important result showed in Fig.~\ref{log3dt0200} is that the information on the
quarkyonic phase is never lost. The full boundary curves constructed by defining
$\rho_s/\rho_{s(\mbox{\tiny vac})}=1/2$ and $\Phi=0.4$ always delimit a phase where the
chiral symmetry is restored and the quarks are still confined. Therefore, it is possible
to represent all the phases and boundaries of strongly matter also with EPNJL model.
Moreover, notice also that the EPNJL model provides the emergence of the quarkyonic phase
only from $\mu_q\sim240$~MeV, for $G_V=0$, differently for the case of the PNJL model of
Fig.~\ref{log3dt0205}, where the quarkyonic phase starts at $\mu_q\sim 130$~MeV.

Finally, we stress here that our method of construction of quark phase diagrams with
all possible boundaries, making the projections of $\rho_s$ and $\Phi$, suggests that the
order parameters should be more correlated each other in order to unambiguously define the
boundaries from the magnitude of $\rho_s$ and $\Phi$. As our results point out, the EPNJL
model, that present such a correlation at any temperature and chemical potential, seems to be a
better candidate to describe the strongly interacting matter phase diagrams than the PNJL model
itself. This result corroborates the lattice QCD calculation that points out to this
correlation at $\mu_q=0$, since it obtains the same temperature transition related to both
order parameters, namely, $T(\mu_q=0)=173\pm 8$~MeV.

\section{Summary and conclusions}
\label{summary}

In this work we have proposed a method of identification of the phases and boundaries
of strongly interacting matter obtained through PNJL model. This method consists in
analyze the magnitude of the order parameters $\rho_s$ and $\Phi$ by projecting their
surfaces in the $T\times\mu_q$ plane. Therefore, it is natural to localize the broken/restored
chiral symmetry and confinement/deconfinement phases, see an example of such a projection for
the RRW06 parametrization in Fig.~\ref{log3d}. The projections also allows the determination of
a particular value of $\rho_s/\rho_{s(\mbox{\tiny vac})}$ and $\Phi$ used to construct the
boundaries in the phase diagram. In the case of RRW06 model, the boundary curves are defined as
those in which $\rho_s/\rho_{s(\mbox{\tiny vac})}=1/2$ and $\Phi=0.4$ represented by the full
curves in Fig.~\ref{log3d-all}. We also compared our boundary curves with those determined
through the peaks of $\partial\rho_s/\partial T$ and $\partial\Phi/\partial T$, frequently used
in the literature, and shown that the quarkyonic phase found by the latter is underestimated
when compared to the found by the former.

The vector repulsive interaction in the PNJL model was other important aspect studied
in our work. We have shown that is not possible to construct two boundaries
in the phase diagram if only peaks of $\partial\Phi/\partial T$ are taken into account. The
double peak structure of $\partial\Phi/\partial T$ in Fig.~\ref{peaks}a is changed to that one
depicted in Fig.~\ref{loggv02}. By using the fixed values of the order parameters from the
aforementioned projections, it is natural to define the boundary curves even for $G_V>0$
cases. It is also clear that the increase of $G_V$ decreases the quarkyonic phase, see
Fig.~\ref{gvevolution}.

Finally, we have investigated the influence of the $T_0$ parameter of the Polyakov potential
RRW06 given in Eq.~(\ref{rrw06}). If we keep the value of $T_0=270$~MeV in the original version
of the RRW06 parametrization, the boundary curves constructed via peaks criterium or by the
magnitude of the order parameters (projection on $T\times\mu_q$ plane) give a value greater
than $200$~MeV for the transition temperature at $\mu_q=0$, what is not supported by lattice
QCD calculations, that give the result of $173\pm 8$~MeV. The common procedure adopted in the
literature is the rescaling of $T_0$ from $270$~MeV to $190$~MeV, taking the average
values of the transition temperatures associated with the peaks of
$\partial\rho_s/\partial T$ and $\partial\Phi/\partial T$ (not coincident for $T_0$
rescaled). This generates only one boundary curve and does not allow the emergence of a
quarkyonic phase in the SU(2) version of the PNJL model. In our method, based on the
analysis of the projection of the order parameters, the suitable rescaling is change $T_0$
from $270$~MeV to $205$~MeV. This is the value that allows us to construct the two
boundary curves presented in Fig.~\ref{log3dt0205} starting at the same point at $\mu_q=0$
and presenting compatible values for the transition temperature. In this case, the curves
have the values of $0.77$ and $0.44$ associated with $\rho_s/\rho_{s(\mbox{\tiny vac})}$
and $\Phi$, respectively. The region surrounded by the two curves is the quarkyonic
phase.

We also concluded that the value of $0.77$ is not unique to define a boundary
curve of the broken/restored chiral symmetry phases for $T_0=205$~MeV. Such an
ambiguity can be removed if the correlation between the order parameters is increased by
making $G_s\to G_s(\Phi)$. The EPNJL model constructed in this way produces the
projections of $\rho_s$ and $\Phi$ obtained in Fig.~\ref{log3dt0200}. The boundary of the
broken/restored chiral symmetry phases is again uniquely defined by the
$\rho_s/\rho_{s(\mbox{\tiny vac})}=1/2$ curve and present a value of
$T(\mu_q=0)\sim170$~MeV, compatible with the lattice result for this quantity. Also the
$\Phi=0.4$ curve stars at the same point at $\mu_q=0$. From this perspective, our
results show that the EPNJL model describe in an unambiguous way all phases and
boundaries of quark matter, better than the PNJL model. This indicate that the correlation
between the order parameters of the quark phase transition must be strongly correlated, as
the lattice QCD results for the temperature transition at $\mu_q=0$ point out.

As a last remark, we stress here the importance and needed of theoretical studies of the
strongly interacting matter phase diagram, mainly in the region of higher chemical potentials
(compressed matter), since this is a region where new experiments will focus in the near
future, for instance at Facility for Antiproton and Ion Research (FAIR)~\cite{fair}, and at
Joint Institute for Nuclear Researches (JINR)~\cite{nica}.

\section*{Acknowledgements}

The authors acknowledge the partial support from Funda\c c\~ao de Amparo \`a Pesquisa do Estado
de S\~ao Paulo (FAPESP) and Conselho Nacional de Desenvolvimento Cient\'\i fico e
Tecnol\'ogico (CNPq) of Brazil.

\end{document}